\documentclass[11pt,a4paper, preprint,  superscriptaddress]{revtex4}
\usepackage[utf8]{inputenc}
\usepackage{hyperref}
\usepackage{pdflscape}
\usepackage{romannum}
\usepackage{amsmath}
\usepackage{amssymb}
\usepackage{subfigure}
\usepackage{epsfig}
\usepackage{epstopdf} 
\usepackage{amsmath,amsfonts}
\usepackage{graphicx}
\begin{document}
	\title{\Large   Study of the large scale structure through modified gravity theory using statistical mechanics
	}
	\author {Abdul W. Khanday}
	\email{abdulwakeelkhanday@gmail.com}
	\affiliation{{Department of Physics, National Institute of Technology  Srinagar, Jammu and Kashmir -190006, India.}}
	\author {Sudhaker Upadhyay}
	\email{sudhakerupadhyay@gmail.com}
	
	\affiliation{Department of Physics, K. L. S. College, Nawada, Bihar 805110, India}
	\affiliation{Department of Physics, Magadh University, Bodh Gaya,
		Bihar  824234, India}
	\affiliation{Inter-University Centre for Astronomy and Astrophysics (IUCAA) Pune, Maharashtra-411007 }
	\affiliation{School of Physics, Damghan University, Damghan, 3671641167, Iran}

	\author {Prince A. Ganai}  
	\email{princeganai@nitsri.net}
	\affiliation{{Department of Physics, National Institute of Technology  Srinagar, 
			Jammu and Kashmir -190006, India.}}

	\begin{abstract}
		We discuss the galaxy clustering based on thermodynamics and statistical mechanics in the expanding universe in a modified theory of gravity. The modified general relativity (MGR) is developed using the regular line element field to construct a symmetric tensor that represents the energy momentum of the gravitational field.  This in turn provides a modified gravitational potential with terms that represent  dark matter and dark energy effects without actually invoking the two. Based on the modified gravitational potential we calculate the distribution function of the galaxies. We also calculate various thermodynamic equations of state. We make a data analysis of the data  obtained through the SDSS-\Romannum{3} survey  and check the feasibility of the theoretical model of probability distribution of galaxies in the universe.  
	\end{abstract}	
	\maketitle
 	\section{Introduction}
	The large scale distribution of visible matter is mainly influenced by the gravity due to the matter itself. Today we believe that the  weak density gradient in the early universe has evolved to the present day large scale cosmic structure. \\
	 Large scale structure is important to the fundamental understanding of the universe due to its slow evolution with time. The structures we observe today are more or less the fossils of the early conditions in the universe. \\
	  The first structures in the universe formed at a red-shift of 10- 30 in halos of dark matter with masses around , $10^5 - 10^8 $ solar masses~\cite{2}. All this structure was formed on the seeds of small fluctuation of matter density in the primordial density field. Today N-body simulations have confirmed the potential of the perturbations in the  initial density field to grow to the present day observed structure ~\cite{3}.
	\par
	Zwicky, in 1933, while studying the red shift of various galaxy clusters noticed a velocity dispersion with a spread of about $2000 $ Km/s within the Coma cluster. 
	Zwicky for the first time concluded that the velocity of galaxies in Coma cluster is not directly correlated  with the total visible mass. He found this mass to be is less by factors of 200 - 400 than the mass required to provide the necessary gravitational field in the cluster ~\cite{4}. The term "Dark matter" was used for the first time by Zwicky to account for the missing mass.\\
	Today we believe that the ratio of dark matter  in the universe is $\approx 27\%$. although there is an overwhelming observational evidence of the existence of dark matter(DM) from galactic to cluster scales. Yet one of the greatest puzzles of modern particle physics and cosmology is the understanding of Dark Matter. After decades of effort there is no direct clue  to understand the basics properties of Dark matter.    
	Recently,a correlation between radial acceleration and baryonic distribution was reported in~\cite{5}.  This can be a result of a strong correlation between baryonic matter and Dark Matter. The other possible explanation could be applicability of a modified dynamics at large scales, e,g. MoND. 
	
	\par
	The study of galaxy clusters is important as they act as astrophysical laboratories as well as probes for the study of large scale structure of the universe. Theses massive bodies provide environment in which many interesting large scale phenomenon can be studied. The formation and evolution of these structures contain information about the evolution of the universe as a whole. The multi-wavelength observations have been very usefull in the study of different processes going on inside clusters. Radio-wave, X-ray and other spectroscopic techniques have helped to evaluate high temperature phenomenon inside clusters.

	\par
	Many theoretical models models e,g. Kaiser ~\cite{6}, of galaxy clusters  have  focused on different properties of clusters in developing the models.  
	\par

 	Saslaw and   Hamilton in 1984~
	\cite{7} developed a new theory of galaxy clustering  in an expanding universe. This model predicts distribution of  all orders from galaxies to voids corresponding to over-dense and under-dense regions respectively. The theoretical  probability distribution function predicted  for $N$ number of galaxies in  a given volume $V$ is 
	\begin{equation}
	f(N)=e^{-\bar{N}(1-b)-Nb}\frac{\bar{N}(1-b)}{N!}\left[\bar{N}(1-b)+Nb\right]^{N-1},\label{01}
	\end{equation} 
	where $\bar{N} = nV$ is the average number of particles in volume $V$. 
	\par
	It is  now established that the force required to describe the  flat rotation curve of galaxy clusters at larger distances should be greater than predicted by the Newtonian theory of gravity. Similarly, the predictions of GR also does not account for the enhanced force at large scale.  
	In the recent past there have been a number of attempts to account for this enhanced interaction on cosmological scales resulting in numerous modifications to general relativity ~\cite{8} to account for this discrepancy. E.g., in $f(R)$ theory of gravity  an additional term $f(R)$ of Ricci curvature, is added  to account for the enhanced gravitational attraction force on cosmological scales. 
	\par The gravitational force field guides the accumilation of mass at the largest possible scales. The statistical method can be emplyed to study the mass distribution on the largest possible scale
	 ~\cite{9}. 
	In recent past  there has been a tremendous effort to study the effect  of modified  gravity theories  on
	the galaxy clustering through statistical and thermodynamic methods \cite{10,11,12,13,14, 15, 16,16a, 16b,16c, 17,18, 19}. It has been observed that the modified theories of gravity does affect the clustering process through a change in a parameter that is related to the strength of correlation called clustering parameter, originally defined as the ratio of Kinetic to potential energy, $b=K/2W$. E.g., in~\cite{14}  the modified clustering/correlation parameter has been studied as a function of the correction  term. The study has shown an enhanced correlation for an increasing strength of the correction term.
	\par
	The use of a smooth regular line element field to construct a symmetric tensor that represents the energy momentum of the gravitational field provides an extra freedom required to describe dark matter and dark energy without destroying the Lorentz invariance of General Relativity (GR). This also leads to a modified Newtonian potential\cite{20} 
	\begin{equation}
	F=-\frac{GMm}{r^2}-\frac{bmc^2}{r}+\frac{a_0mc^2}{2},\label{eq02}
	\end{equation} 
	 where $b$ and $a_0$ are parameters and  $b>0$.
	 The first term is the usual Newtonian attractive force. The second term is also attractive and represents the gravitational attraction due to Dark matter while the third term is repulsive and represents Dark energy.
	\par
	Although the gravitational galaxy clustering has been studied extensively under various theories of gravity, the study under  the modified gravity described in(\cite{20} ) has not been studied. In this paper we try to bridge this existing gap. We will  also make a comparison of the theory and the available data.\\ 
	The structure of the paper is as follows. First we construct the partition function (section II) for the system of galaxies. In section \Romannum{3} we derive the various thermodynamic equations of state that are meaning full here, e,g. Helmholtz free energy, specific entropy, pressure, chemical potential among others. In next section (\Romannum{4}) we make a comparison of the clustering parameter with increasing radial distance for the Newtonian and MGR gravity laws. In section (\Romannum{5}) we study the effects of this new  force form   on  the  statistical distribution of the galaxy clusters. Finally, in section (\Romannum{6})  we analyze the model by fitting data to theory.
	In section(\Romannum{7}), we make a discussion and conclusion.
	
	\section{THE PARTITION FUNCTION}
	Here we develop the partition function from ab-initio of the system of particles interacting  pairwise through the modified  gravitational interaction. We assume our particles to be in co-moving ensemble of cells in the expanding background. Let the volume of the cells be $V$ containg an average number density $\bar{N}$ of particles. The partition function (canonical) for this system  is: 
	\begin{equation}
	Q= \frac{1}{N!} \int...\int e^{-H/T} dp_1....dp_Ndr_1...dr...N,
	\end{equation} 
	where $H$ represents the total energy of the system. Here the factor $\lambda$ is a normalization constant and $N!$ takes care of distinguishability of system of particles. The Integral can be further simplified to
	\begin{equation}
Q=	\frac{Z	}{N! \Lambda^{3N}}\label{03}
 	\end{equation}
	where
	\begin{equation}
	 \Lambda=\frac{h}{(2\pi m T)^{1/2}}\notag
	\end{equation}
	and 
	\begin{equation}
	Z_N= \int...\int e^{U/T} dr_1...dr_N.\label{5}
	\end{equation}
	   Here we have set the Boltzmann's constant equal to unity. The configuration integral, equation (\ref{5}) can be written as,
	\begin{equation}
	Z_{N}(T,V)=\displaystyle \int...\int  \exp  \left[-T^{-1}U(r_{1},r_{2},...r_{N})\right] d^{3N}r.\label{eq4}
	\end{equation}
	The gravitational potential energy
	function $U(r_{1},r_{2},...r_{N})$ is merely the sum of  the potentials of   the all pairs of particles. That is,
	\begin{equation}
	U(r_{1},r_{2},...r_{N})=\sum_{1\le i\le j\le N}^{}u(r_{ij})\label{7}
	\end{equation}
	where $u(r_{ij})$ is the pairwise potential. Here we introduce a function $f_{ij}$ defined by
	\begin{equation}
	f_{ij}= e^{-u(r_{ij})}-1\label{eq8}.
	\end{equation}
	Rewriting equation (\ref{eq8}) in the following form 
	\begin{equation}
	e^{-U/KT}=e^{\sum u(r_{ij}/T)}=\prod e^{-u(r_{ij})/T}=\prod_{1\le i\le j\le N}(1+f_{ij}).\label{eq9}
	\end{equation}
	The $f$ functions takes care of the pair-wise interaction of the system particles and in the absence of any interaction it reduces to zero.

	 In terms of this two-point function  equation (\ref{eq4}) now  takes the following  form:  
	\begin{equation}
	Z_{N}(T,V)=\displaystyle\int...\int(1+f_{12})(1+f_{13})(1+f_{23})(1+f_{14})...(1+f_{N-1,N})d^{3}r_{1}d^{3}r_{2}...d^{3}r_{N}.\label{eq10}
	\end{equation}
	By the incorporation of the explicit for of the modified gravitational potential function (eqn. (\ref{eq02})), the two point function $f$ takes the form
	
	\begin{equation}
	\large	f_{ij}=\frac{A_1}{(r_{ij}^2+\epsilon^2)^{1/2}T}+A_2 \log(r)+A_3 r.\label{eq11},
	\end{equation}
	where 
	\begin{align*}
	A_1 &=  \frac{-GMm}{R}\\
	A_2 &= -bmc^2 \\
	A_3 &= a_0mc^2.
	\end{align*}
	Here we incorporated a parameter called softening parameter $\epsilon$, with range  $\epsilon$ is $0.01\le \epsilon \le 0.05$, to avoid the divergence of the Hamiltonian at the origin i.e, $r_{ij}=0$. The softening parameter is not needed in the 2nd and 3rd term as there is no chance of divergence of the function.

	By substitution of  equation (\ref{eq11}) in equation (\ref{eq10}), the values of $Q_N$ for different values  of $N$ can be calculated. E.g, for $N = 1$, we have
	\begin{equation}
	Z_{1}(T,V)= V.\notag
	\end{equation}
	For $N=2$,  we evaluate the integral by fixing  $r_1$ and evaluating over all the other particles.  In this way the integral simplifies to 
	\begin{equation}
	Z_{2}(T,V)=V^{2}\bigg[1+ A_1\alpha_1+A_2\alpha_2+A_3\alpha_3\bigg],\notag
	\end{equation}
	where
\begin{align*}
\alpha_1 &= \Big[\frac{3}{2R}+\frac{3}{2R^3}\epsilon^2\log\frac{\epsilon/R}{1+\sqrt{1+(\epsilon/R)^2}}\Big]\\
\alpha_2 &= \Big[\frac{-1}{3}+\log(R)\Big]\\
\alpha_3 &= \Big[\frac{3A_3R}{4}\Big]
\end{align*}
	Similarly  for the higher values of $N$ i.e, $N = 3, 4, 5, ..., N$, the value of $Z_N$  can be obtained. E.g, for $N = 3$, we have 
	\begin{equation}
	Z_{3}(T,V)=V^{3}\bigg[1+ A_1\alpha_1+A_2\alpha_2+A_3\alpha_3\bigg]^{2}.\notag
	\end{equation}
	In general we can write  the above equation for any   number of  particles $N$ as, 
	\begin{equation}
	Z_{N}(T,V)=V^{N}\bigg[1+ A_1\alpha_1+A_2\alpha_2+A_3\alpha_3\bigg]^{N-1}.\label{eq12}
	\end{equation}
	Finally,  on substituting equation (\ref{eq12}) into equation (\ref{eq4}),  the general form of the partition function for a gravitating system of $N$ pairwise interacting particles (galaxies)  is
	\begin{equation}
	Q_{N}(T,V)=\frac{1}{N!}\bigg(\frac{2\pi mT}{\Lambda^{2}}\bigg)^{\frac{3N}{2}} V^{N}\bigg[1+A_1\alpha_1+A_2\alpha_2+A_3\alpha_3\bigg]^{N-1}.\label{eq13}
	\end{equation}
	Equation (\ref{eq13}) is the standard form of canonical partition function for a system of $N$ particles interacting through the new gravity law . All the effects of the modification are incorporated in the terms $\alpha_1$ and $\alpha_2$.

	\section{ THERMODYNAMIC EQUATIONS OF STATE}
	Here we study the effect of the corrected Hamiltonian on the various thermodynamic equations of state. Utilizing the gravitational partition function along with the standard relations, we can compute compute the equations of state. For instance, the Helmholtz free energy can be derived from the standard relation $F=-T\ln Q_{N}(T,V)$. We substitute the value of $Q_N$ from equation \ref{eq13} and obtain the modified Helmholtz free energy of the system as; 
	
	\begin{equation}
	\begin{split}
	F&=NT\ln\bigg(\frac{N}{V}T^{\frac{-3}{2}}\bigg)-NT-\frac{3}{2}NT\ln\bigg(\frac{2\pi mT}{\Lambda^{2}}\bigg)-NT\ln [1+A_1\alpha_1+A_2\alpha_2+A_3\alpha_3]\\
	&=NT\ln\bigg(\frac{N}{V}T^{\frac{-3}{2}}\bigg)-NT-\frac{3}{2}NT\ln\bigg(\frac{2\pi mT}{\Lambda^{2}}\bigg)+NT\ln [1-b_m].
	\label{eq14}
	\end{split}
	\end{equation}
	 In the above equation (\ref{eq14}), the parameter $b_m$ is defined as
	\begin{equation}
	b_m=\frac{A_1\alpha_1+A_2\alpha_2+A_3\alpha_3}{ 1+A_1\alpha_1+A_2\alpha_2+A_3\alpha_3 }.\notag
	\end{equation} 
	This $b_m$ is the  new correlation parameter  which estimates the strength of clustering that in turn governs the  time evolution of the galaxy cluster. The parameter $b_m$ can take values between $0$ and $1$.
	
	\begin{figure}[h!]
		\centering
		\includegraphics[width=8 cm, height=6 cm]{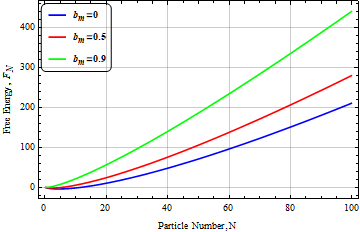}
		\caption{ The plot shows variation of the free energy as a function of particle number for different values of the $b_m$.}\label{fig1}
	\end{figure}
	Similarly other thermodynamic quantities can be estimated,  e.g,  the entropy of the system  can be calculated utilizing the standard relation, $S=-\bigg(\frac{\partial F}{\partial T}\bigg)_{V,N}$. Substituting equation(\ref{eq14}) the entropy of the system of galaxies takes the  form:
	\begin{equation}
	S=N\ln\biggl(\frac{V}{N}T^{3/2}\biggl) +N \ln [1+A_1\alpha_1+A_2\alpha_2+A_3\alpha_3 ] -3N\frac{A_1\alpha_1+A_2\alpha_2+A_3\alpha_3}{ 1+A_1\alpha_1+A_2\alpha_2+A_3\alpha_3} + \frac{5}{2}N +\frac{3}{2}N \ln\bigg(\frac{2\pi m }{\lambda^{2}}\bigg).\label{eq15}
	\end{equation}
	\par

	 Specific entropy  per particle of the system corresponds to equation (\ref{eq15}) can be written as 
	\begin{equation}
	\frac{S}{N}=\ln\biggl(\frac{V}{N}T^{3/2}\biggl)-\ln[1-b_m]-3b_m
	+\frac{5}{2}+\frac{3}{2}\ln \frac{2\pi m}{\lambda^2},\label{eq16}
	\end{equation}
	\begin{figure}[h!]
		\centering
		\includegraphics[width=8 cm, height=6 cm]{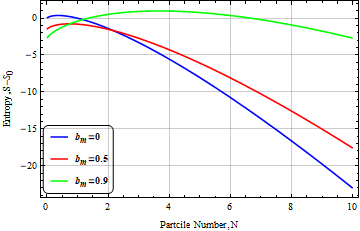}
		\caption{ The  variation of entropy ($S-S_0$) with a changing  particle number for different values of the parameter $b_m$.}\label{fig2}
	\end{figure}

	Using the basic definition, $ U = F + T S$, the internal energy of the system can be calculated. Plunging in the value of  free energy (\ref{eq15})  and entropy  (\ref{eq16}), the internal energy in terms of the new clustering parameter can be written as
	\begin{eqnarray}
	U&=&\frac{3}{2}NT\bigg[1-2\frac{A_1\alpha_1+A_2\alpha_2+A_3\alpha_3}{ 1+A_1\alpha_1+A_2\alpha_2+A_3\alpha_3 } \bigg]\nonumber\\
	&=&\frac{3}{2}NT\left[1-2b_m\right].\label{eq17}
	\end{eqnarray}
	\begin{figure}[h!]
		\centering
		\includegraphics[width=8 cm, height=6 cm]{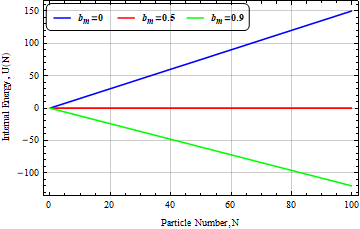}
		\caption{ The plot shows the   behavior of internal energy $U$ with changing particle number for different values of clustering parameter $b_m$.}\label{fig3}
	\end{figure}
	The graphical representation of the effect of the modified clustering parameter on the internal energy function for  the system of galaxies can be visualized in Fig. \ref{fig3}. 
	
	The pressure caused by the particles inside the system can be calculated  utilizing the fundamental relation $P=-\bigg(\frac{\partial F}{\partial V}\bigg)_{T,N}$ as follows;
	
	\begin{align}
	P&=\frac{NT}{V} \left[1-\frac{A_1\alpha_1+A_2\alpha_2+A_3\alpha_3}{ 1+A_1\alpha_1+A_2\alpha_2+A_3\alpha_3 } \right],
	\nonumber\\ 
	&=\frac{NT}{V}\left[1-b_m\right].\label{eq18}
	\end{align}
	
	\begin{figure}[h!]
		\centering
		\includegraphics[width=8 cm, height=6 cm]{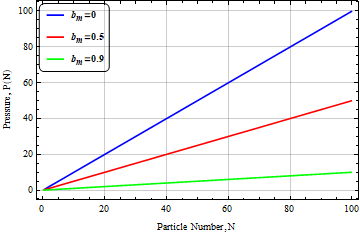}
		\caption{ The change in the pressure created by the particles and gas in the system with increasing particle number  for different values of clustering parameter $b_m$.}\label{fig4}
	\end{figure}
	\par
\begin{figure}[h!]
		\centering
		\includegraphics[width=8 cm, height=6 cm]{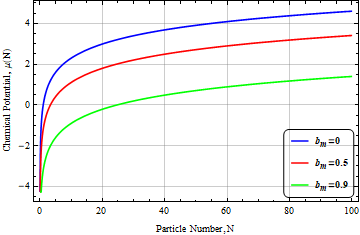}
		\caption{ The variation  of  the chemical potential $\mu (N)$ of the system  with a changing particle number for three values of clustering parameter $b_m$.} \label{fig5}
	\end{figure}

 
	In the similar fashion ,the chemical potential can be calculated  using the fundamental relation, 
	$\mu=\bigg(\frac{\partial F}{\partial N}\bigg)_{T,V}$ as
	\begin{eqnarray}
	\mu &=&T\bigg(\ln\frac{N}{V}T^{-\frac{3}{2}}\bigg)+T\ln\bigg[1-\frac{A_1\alpha_1+A_2\alpha_2+A_3\alpha_3}{ 1+A_1\alpha_1+A_2\alpha_2+A_3\alpha_3 }\bigg]-T\frac{A_1\alpha_1+A_2\alpha_2+A_3\alpha_3}{ 1+A_1\alpha_1+A_2\alpha_2+A_3\alpha_3 }-\frac{3}{2}T\ln\bigg(\frac{2\pi m}{\lambda^{2}}\bigg),\nonumber\\
	&=&T\left(\ln \frac{N}{V}T^{-3/2}\right)+T\ln \left[1-b_m\right]-Tb_m
	-\frac{3}{2}T\ln\left(\frac{2\pi M}{\lambda^2}\right).\label{eq19}
	\end{eqnarray} 
	Figure (\ref{fig5}) shows graphical representation of the change in the chemical potential with an increase in the particle number.	
	\section{A study of standard clustering parameter in comparison to modified clustering parameter}
		Here we discuss the effect of the correction to the gravity theory on the clustering parameter that estimates the strength of the interaction which in turn can tell us about the time-scale of clustering and hence that of structure formation.
	\begin{figure}[h!]
		\centering
		\includegraphics[width=8 cm, height=7 cm]{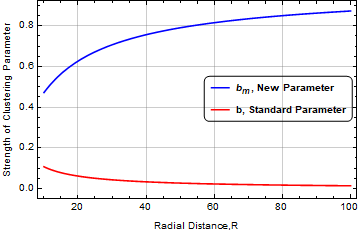}
		\caption{ Graphical comparison of the standard clustering parameter $b$ with the new parameter $b_m$}\label{fig6}
	\end{figure}
It is evident from the graph that the modified clustering parameter is stronger then the standard one. This is a direct consequence of the increased strength of the potential in the modified gravity model.\\
	\section{General distribution function}
	The general distribution function $f(N)$, which gives the distribution of the over-densities and voids in  fixed volume cells, can be developed in the modified gravity model utilizing the partition function developed in eqn. \ref{eq13} . We let the particle number to change, possibly through mergers and boundary crossings, characterized by the chemical potential of the system \ref{eq19}. The distribution function can be developed as;  
	\begin{equation}
	Q_G(T,V,z)=\sum_{N=0}^{\infty}\exp \left(\frac{N\mu}{T}Q_N(T,V)\right).\label{eq20}
	\end{equation} 
	The probability of finding $N$ particles in a cell of volume $V$ in a  grand ensemble is given by
	\begin{eqnarray}
	F(N)&=&\sum_{i=0}^{N}\frac{\exp\frac{N\mu}{T}\exp\frac{-U}{T}}{Q_G(T,V,z)},\nonumber\\
	&=&\frac{\exp\frac{N\mu}{T}Q_N(T,V)}{Q_G(T,V,z)}.\label{eq21}
	\end{eqnarray} 
	Here,  $z=\exp\frac{\mu}{T}$ (fugacity) determines the activity of the system towards particle number change.  From the  equation (\ref{eq21}), we can easily determine the general form of distribution function for the system of galaxies. Using the relation for the partition function (\ref{eq20}) along with the  chemical potential  equation \ref{eq12}  the distribution function takes the  following form:
	\begin{equation}
	F(N)= \frac{\bar{N}}{N!}(1-b_m)\bigg[\bar{N}(1-b_m)+Nb_m\bigg]^{N-1} \exp-Nb_m-\bar{N}(1-b_m).\label{eq22}
	\end{equation}
	The distribution function \ref{eq22} has a Poisson structure with additional terms that incorporate the effect of the correction. The graphical representation of the distribution with an without correction parameter is shown in the figure 
	
	\begin{figure}[h!]
		\centering
		\includegraphics[width=8 cm, height=6 cm]{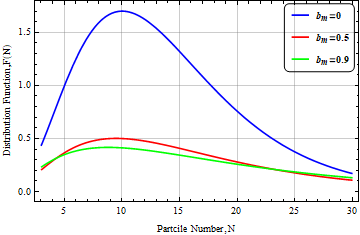}
		\caption{ The plot shows the change in the distribution function $F(N)$ with increasing   particle number, $N$ for different values of $b_m$.}
		\label{fig7}\end{figure} 
	\section{observational data}
\begin{figure}[h!]
	\centering
	\includegraphics[width=8 cm, height=6 cm]{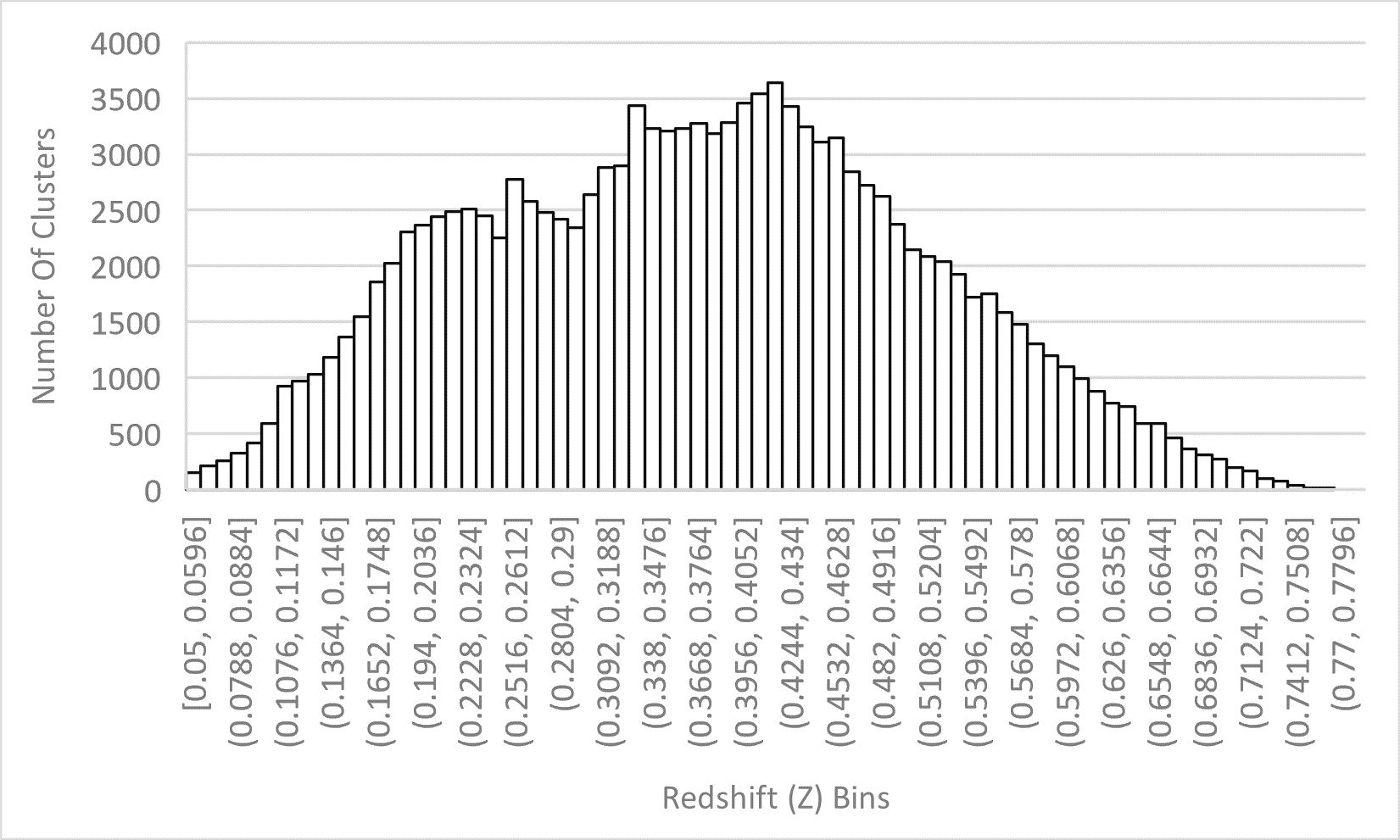}
	\caption{  The red-shift distribution ($z$) of the 132,684 identified clusters in the sky survey SDSS-III}
	\label{fig8}\end{figure}

Here we test the feasibility of the effect of  modified gravity theory on galaxy clustering by choosing a galaxy/cluster catalog and make a comparison of our developed model with the data. The cluster catalog we choose to test the model is the \cite{21} which contains 132,684 cluster in the red-shift range of 0.05-0.8. The catalog uses the observational data from the Sloan Digital Sky survey \Romannum{3} (SDSS-\Romannum{3}). The catalog provides details of the radius within which the mean density of a cluster is $\approx$ 200  ($r_{200}$)\footnote{The $r_{200}$  is represented by $R$ in graphs(\ref{fig11}) and in table(\ref{tab:results}) and it is different from the constant $R$ mentioned in section(\Romannum{2}). } along with number of clusters in $r_{200}$ i.e., $N_{200}$. Figure (\ref{fig8}) shows the number of galaxy clusters observed in different red-shift bins.
\par 
We bin our data  first  on the basis of red-shift ($z$) and then on the basis of radius ($r_{200}$). We fix the length of the bins in terms of  red-shift $(z)$ and radius $r_{200}$  viz $\Delta z = 0.15$ and $\Delta r_{200} = 0.25 Mpc $. Here we determine the cells by physical boundaries i.e., radius $r_{200}$. We derive the the probability distribution functions for each cluster by  counting the galaxy number in each cluster/cell. We used the Application Programming Interface (API) Scipy.optimize.curve$\_$fit of SciPy python library to fit our model and obtain the optimized parameter values. This module provides the control and flexibility to define the form of model curve, where optimization is used to locate the optmal values values of the parameters of the model function. The optimized value of the clustering parameter $b_m$ for different radii and red-shift ranges is enlisted in the table (\Romannum{1}) after fitting equation(\ref{eq22}).

\par

From the plots we note that the fit is best for larger values of radii i.e., $1.15<R<1.40  Mpc$, fig(\ref{fig11}, (c,f,i)). For low $R$ values the model fits well with clusters having larger number of galaxies, $N>30$. In figure(\ref{fig12}) we have also plotted the Ra and DEC sky distribution of the galaxy clusters in the sky. 	
		\begin{figure}[h!]
		\centering
		\subfigure[]{\includegraphics[width=0.40\textwidth]{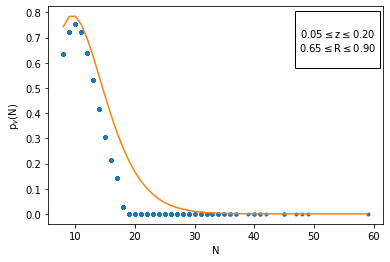}} 
		\subfigure[]{\includegraphics[width=0.40\textwidth]{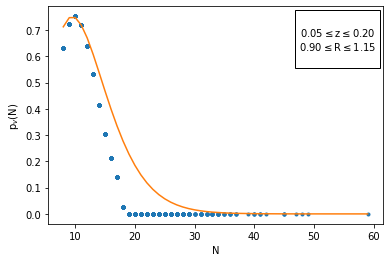}}\\
		\subfigure[]{\includegraphics[width=0.40\textwidth]{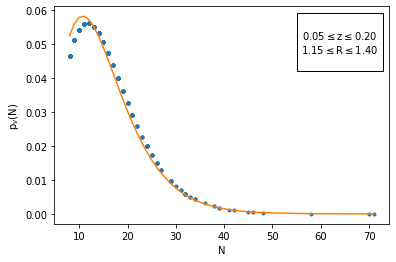}}
		\subfigure[]{\includegraphics[width=0.40\textwidth]{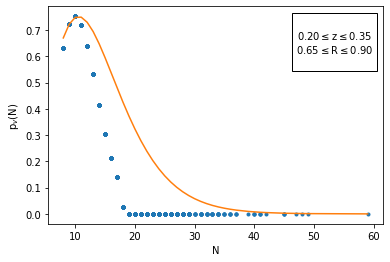}}\\
		\subfigure[]{\includegraphics[width=0.40\textwidth]{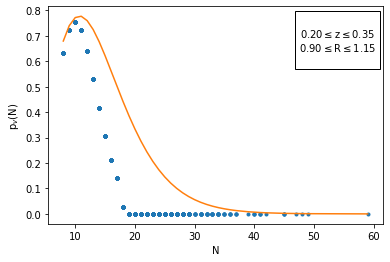}} 
		\subfigure[]{\includegraphics[width=0.40\textwidth]{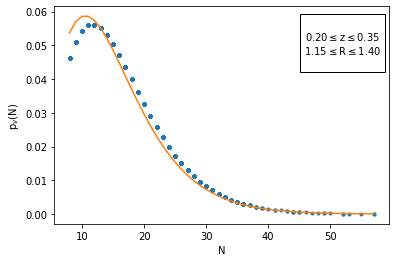}}
	\end{figure}
\begin{figure} 
		\subfigure[]{\includegraphics[width=0.40\textwidth]{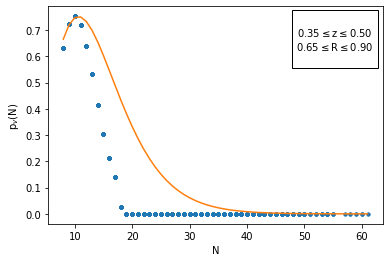}}
		\subfigure[]{\includegraphics[width=0.40\textwidth]{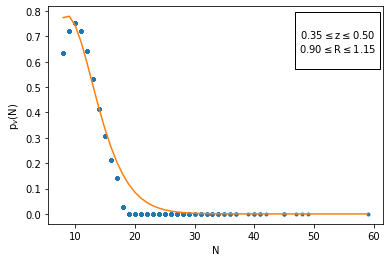}}\\
		\subfigure[]{\includegraphics[width=0.40\textwidth]{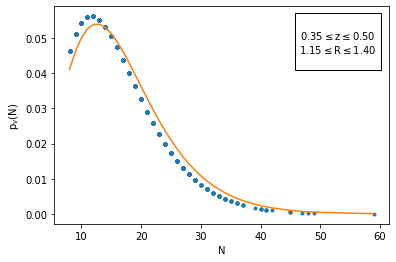}} 
		
		\caption{Probability distribution of galaxy clusters in various red-shift ranges}
		\label{fig11}
	\end{figure}
	\begin{figure}
		\centering
		\subfigure[]{\includegraphics[width=0.40\textwidth]{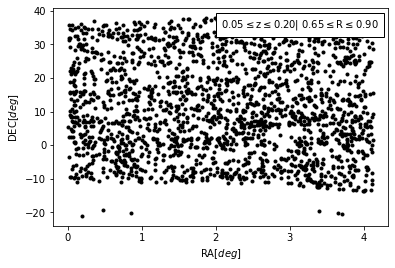}} 
		\subfigure[]{\includegraphics[width=0.40\textwidth]{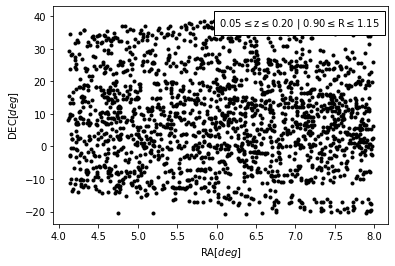}} \\
		\subfigure[]{\includegraphics[width=0.40\textwidth]{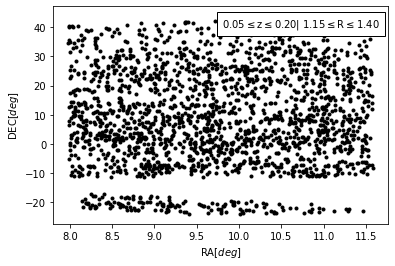}}
		\subfigure[]{\includegraphics[width=0.40\textwidth]{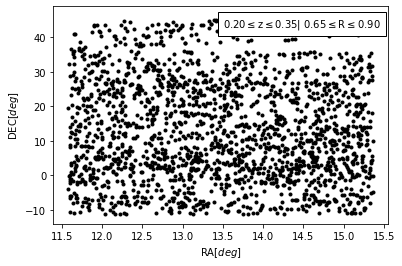}}\\
		\subfigure[]{\includegraphics[width=0.40\textwidth]{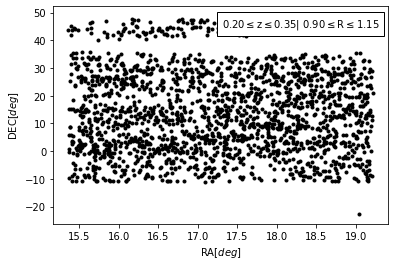}} 
		\subfigure[]{\includegraphics[width=0.40\textwidth]{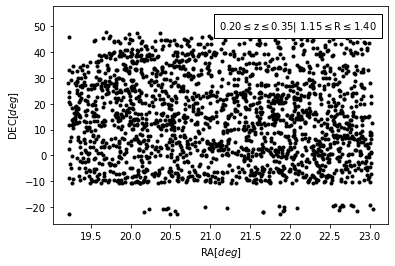}} 
	\end{figure}
\begin{figure}
		\subfigure[]{\includegraphics[width=0.40\textwidth]{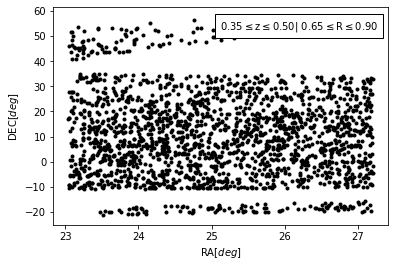}}
		\subfigure[]{\includegraphics[width=0.40\textwidth]{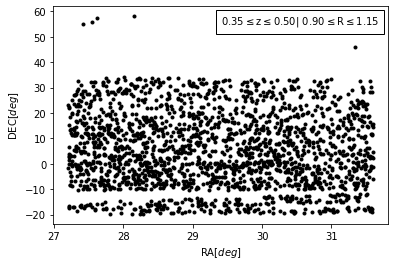}}\\
		\subfigure[]{\includegraphics[width=0.40\textwidth]{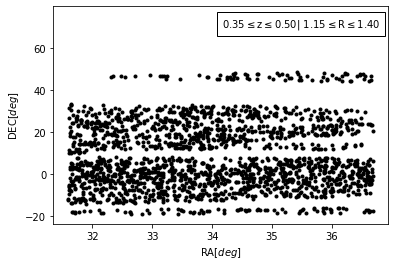}} 
		
		\caption{Sky distribution of clusters shown in various RA(deg) and DEC(deg) coordinates}
		\label{fig12}
	\end{figure}
\begin{table*}
\begin{minipage}{\textwidth}
	\caption{ The table shows the values of different parameters  after the model is fit to the data.  Data is binned  first in redshift ranges and then in radius ranges. $N_cl$, $N_G$  represents the number of clusters and number of galaxies respectively.}\label{tab:results}
	\centering
\resizebox*{\textwidth}{!}{
	\begin{tabular}{|c|cccc|cccc|cccc|}
		\multicolumn{13}{c}{} \\
		\hline
		& \multicolumn{4}{c|}{$0.05<z<0.20 Mpc$} & \multicolumn{4}{c|}{$0.20<z<0.35 Mpc$} & \multicolumn{4}{c|}{$0.35<z<0.50 Mpc$} \\
		$R$ & $N_{cl}$& $N_G$ & $\bar{N}$ & $b_n$ &$N_{cl}$&$N_G$& $\bar{N}$ & $b_n$  
		&$N_{cl}$& $N_{G}$ & $\bar{N}$ & $b_n$   \\
		\hline
		$[0.65,0.90]$ &$382$&  $6272$ & $16.42$ & $0.30$  & $13458$ &$23115$&
		$17.14$ & $0.37$ & $6575$ &$108309$& $16.47$ & $0.0.37$   \\
		$[0.90,1.15]$ &  $755$ &$11618$& $15.39$ & $0.32$  & $21810$ &$329928$&
		$15.13$ & $0.31$ & $26464$ & $370360$ & $13.99$ &
		$4.51$ \\
		$[1.15,1.40]$ &  $247$ &$4389$& $17.77$ & $0.40$  & $4706$ &$69822$&
		$14.84$ & $0.51$  & $4509$ &$64213$& $14.24$ & $0.32$  \\	
		\hline	
\end{tabular}}\label{table}
\end{minipage}
\end{table*}

	\section{ discussion and Conclusion}
	
	In this paper, we have studied the galaxy clustering under a modified theory of gravity, motivated by the inclusion of a smooth regular line element field to construct a symmetric tensor, assuming that the system of galaxies is in quasi-equilibrium state.
	First we calculated the gravitational partition function using the modified gravitational potential. Utilizing the partition function  we also calculated various equations of state viz free energy, entropy, pressure among others. We also analyzed the behavior of the so calculated equations of state and utilized these thermodynamic potentials  to make a comparison between the Newtonian and modified theory of gravity using the equations of state.	
	We could see that the modified gravitational potential has a considerable effect on the various equations of state. E.g., the chemical potential($\mu(N)$) has reduced considerably by the inclusion of the correction terms which implies that particle number within the system does not change much.
	 We  also observed that the clustering parameter  increase in strength for increased value of the  correction parameter. The changing clustering parameter has a direct effect on the time scale of clustering.
	 We also made a comparison of the probability distribution of the galaxies with the observed data obtain from SDSS-\Romannum{3}. In the bins in range $1.15<R<1.40$ the fit is very close, but in bins in ranges $0.65<R<0.90$ and $0.90<R<1.15$ the model fits for the  clusters with large populations ($>30$) but for less populated clusters the fit is not very perfect($<30$).

\end{document}